\documentclass[acmsmall,colorlinks,bookmarksnumbered,unicode]{acmart}

\AtBeginDocument{%
  \providecommand\BibTeX{{%
    \normalfont B\kern-0.5em{\scshape i\kern-0.25em b}\kern-0.8em\TeX}}}

\usepackage[compatibility=false]{caption}

\newcommand{\be}{\begin{equation}}
\newcommand{\ee}{\end{equation}}
\newcommand{\ba}{\begin{eqnarray}}
\newcommand{\ea}{\end{eqnarray}}
\newcommand{\bra}[1]{\langle {#1} |}
\newcommand{\ket}[1]{| {#1} \rangle}
\newcommand{\expect}[1]{\langle {#1} \rangle}

\newcommand{\QS}{Q\#}

\begin{document}

\title{QCOR: A Language Extension Specification for the Heterogeneous Quantum-Classical Model of Computation}

\author{Tiffany M. Mintz}
\email{mintztm@ornl.gov}
\affiliation{%
  \department{Computer Science and Mathematics}
  \institution{Oak Ridge National Laboratory}
  \streetaddress{1 Bethel Valley Road}
  \city{Oak Ridge}
  \state{Tennessee}
  \postcode{37831}
}

\author{Alexander J. McCaskey}
\email{mccaskeyaj@ornl.gov}
\affiliation{%
   \department{Computer Science and Mathematics}
  \institution{Oak Ridge National Laboratory}
  \streetaddress{1 Bethel Valley Road}
  \city{Oak Ridge}
  \state{Tennessee}
  \postcode{37831}
}

\author{Eugene F. Dumitrescu}
\email{dumitrescuef@ornl.gov}
\affiliation{%
  \department{Computational Sciences and Engineering}
  \institution{Oak Ridge National Laboratory}
  \streetaddress{1 Bethel Valley Road}
  \city{Oak Ridge}
  \state{Tennessee}
  \postcode{37831}
}

\author{Shirley V. Moore}
\email{mooresv@ornl.gov}
\affiliation{%
  \department{Computer Science and Mathematics}
  \institution{Oak Ridge National Laboratory}
  \streetaddress{1 Bethel Valley Road}
  \city{Oak Ridge}
  \state{Tennessee}
  \postcode{37831}
}

\author{Sarah Powers}
\email{powersss@ornl.gov}
\affiliation{%
  \department{Computer Science and Mathematics}
  \institution{Oak Ridge National Laboratory}
  \streetaddress{1 Bethel Valley Road}
  \city{Oak Ridge}
  \state{Tennessee}
  \postcode{37831}
}

\author{Pavel Lougovski}
\email{lougovskip@ornl.gov}
\affiliation{%
  \department{Computational Sciences and Engineering}
  \institution{Oak Ridge National Laboratory}
  \streetaddress{1 Bethel Valley Road}
  \city{Oak Ridge}
  \state{Tennessee}
  \postcode{37831}
}

\renewcommand{\shortauthors}{Mintz, et al.}

\begin{abstract}
  Quantum computing is an emerging computational paradigm that leverages the laws of quantum mechanics to perform elementary logic operations. Existing programming models for quantum computing were designed with fault-tolerant hardware in mind, envisioning standalone applications. However, near-term quantum computers are susceptible to noise which limits their standalone utility. To better leverage limited computational strengths of noisy quantum devices, hybrid algorithms have been suggested whereby quantum computers are used in tandem with their classical counterparts in a heterogeneous fashion. This {\it modus operandi} calls out for a programming model and a high-level programming language that natively and seamlessly supports heterogeneous quantum-classical hardware architectures in a single-source-code paradigm. Motivated by the lack of such a model, we introduce a language extension specification, called QCOR, that enables single-source quantum-classical programming. Programs written using the QCOR library and directives based language extensions can be compiled to produce functional hybrid binary executables. After defining the QCOR's programming model, memory model, and execution model, we discuss how QCOR enables variational, iterative, and feed forward quantum computing. QCOR approaches quantum-classical computation in a hardware-agnostic heterogeneous fashion and strives to build on best practices of high performance computing (HPC). The high level of abstraction in the developed language is intended to accelerate the adoption of quantum computing by researchers familiar with classical HPC. \let\thefootnote\relax\footnote{This manuscript has been authored by UT-Battelle, LLC under Contract No. DE-AC05-00OR22725 with the U.S. Department of Energy. The United States Government retains and the publisher, by accepting the article for publication, acknowledges that the United States Government retains a non-exclusive, paid-up, irrevocable, world-wide license to publish or reproduce the published form of this manuscript, or allow others to do so, for United States Government purposes. The Department of Energy will provide public access to these results of federally sponsored research in accordance with the DOE Public Access Plan. (http://energy.gov/downloads/doe-public-access-plan)} 
\end{abstract}

\ccsdesc[500]{Computing methodologies~Parallel programming languages}
\ccsdesc[300]{Computing methodologies~Distributed programming languages}
\ccsdesc[300]{Theory of computation~Object oriented constructs}

\keywords{quantum computing, heterogeneous computing, quantum programming, programming model specification, NISQ}

\maketitle

\section{Introduction}\label{sec:intro}
Quantum computing (QC) holds the potential to solve computational challenges beyond the reach of exascale classical HPC. 
To reap the full benefits of QC, fault tolerant quantum hardware with thousands of logical qubits is required. Such hardware will allow execution of complex quantum algorithms with the user experience similar to how classical algorithms are executed on present day computers. The transition from classical to quantum computing will not be straightforward. Classical algorithms will need to be significantly overhauled to conform to the QC paradigm. This need represents a significant barrier for adoption of QC by the research community. Wide adoption is further complicated by the challenge of developing scientific applications that incorporate quantum expressions to return accurate results when executed on noisy intermediate-scale quantum (NISQ) devices~\cite{preskillNISQ2018} comprised of dozens to hundreds of imperfect qubits. The resulting errors severely constrain the types of quantum algorithms that can be implemented in the absence of fully fault-tolerant resources~\cite{moll2017,perdomo2017}. The hybrid quantum-classical computation (HQCC) model is expected to be one of the most successful ways of using quantum computing in the NISQ era. As researchers develop applications that scale beyond a few qubits, a comprehensive community standard for HQCC programming, compilation, and execution is needed.  

\begin{figure*}[ht]
 \centering
  \includegraphics[width=5in,height=3.5in]{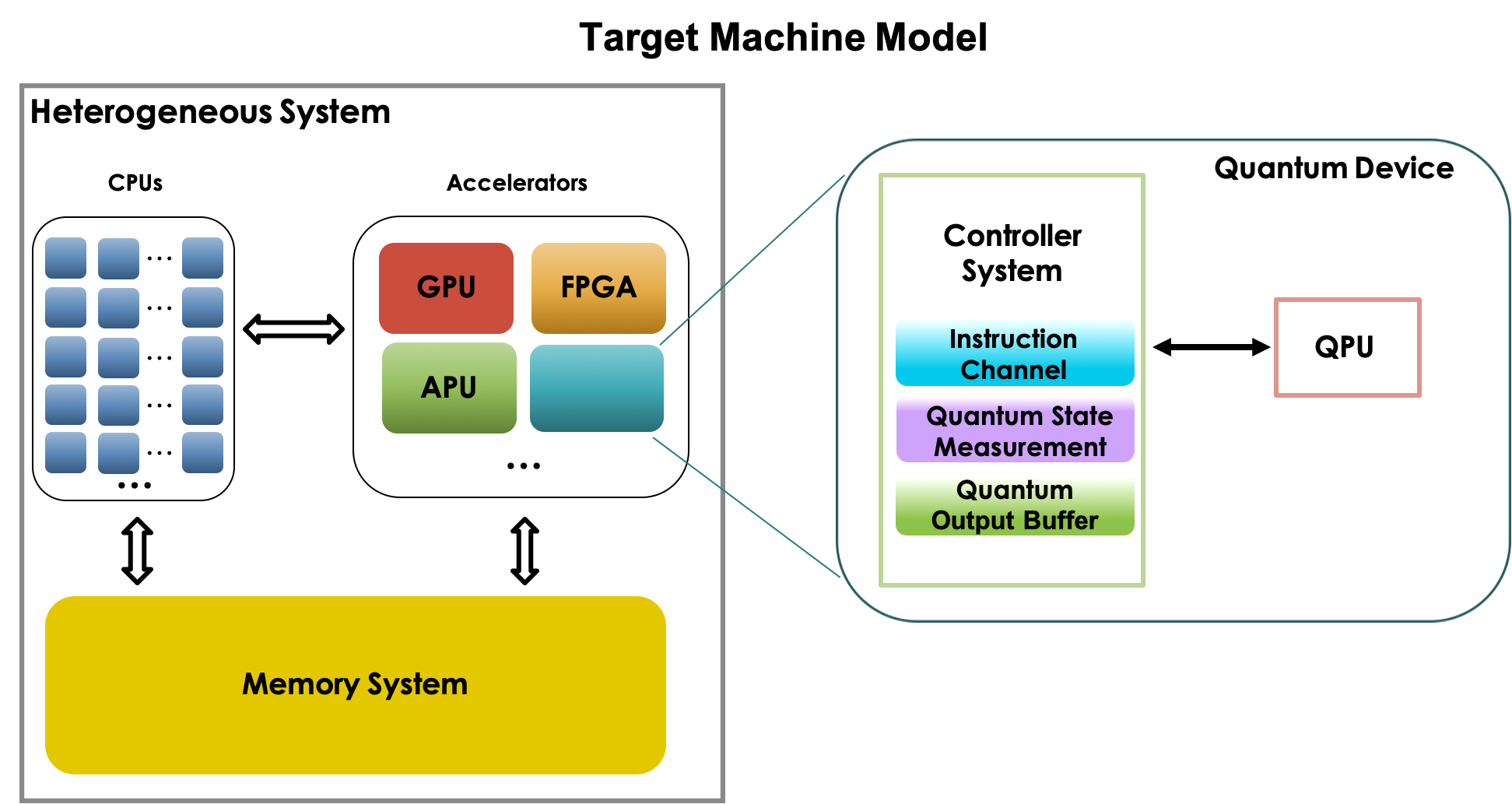}
  \caption{Illustration of the target extreme scale, heterogeneous machine model}
  \label{fig:mach_model}
\end{figure*}

Recent work in the development of quantum computing hardware has resulted in a number of available, open-source frameworks for quantum programming, compilation, and execution. For example, several vendor-specific, high-level Pythonic frameworks have been released \cite{McKay2018, forest, cirq}. These approaches provide circuit construction data structures and enable remote execution of quantum programs via standard Representational State Transfer (REST) application programming interfaces (APIs), thereby promoting a loosely coupled, heterogeneous computational model. This approach has enabled small-scale experimentation, but it becomes unfavorable as applications are developed that utilize large scale classical resources. In order to support future large scale applications of quantum computing, we should consider computational and programming models that best fit the target heterogeneous system architecture. The target systems for quantum computing are extreme scale, heterogeneous, high performance computing (HPC) systems that incorporate quantum devices as co-processors for more traditional, classical processors. Figure \ref{fig:mach_model} shows the target machine model for future systems that will incorporate quantum devices. Writing applications for such a system requires high-level APIs that are interoperable with languages like C and C\texttt{++}. These high-level interfaces should ideally match the level of abstraction and expressiveness of C and C\texttt{++} and offer seamless integration of quantum kernels into a C or C\texttt{++} application.

In this work, we present QCOR, a library and directives based quantum-classical programming language extension that is motivated by the HQCC paradigm, with a focus on iterative and variational algorithmic approaches. 
QCOR's inherently heterogeneous model of computation adopts best programming practices from HPC, enabling QCOR to implement both library calls and directive expressions, along with complex data types and data structures, as part of the language. We have designed the QCOR specification as a language extension for standard classical languages like C and C\texttt{++} to provide generic high-level expressions that can be used to construct quantum algorithms within a single-source, asynchronous approach. 

This work is structured as follows. In Sec.~\ref{sec:existingapproaches} we review existing programming approaches. In Sec.~\ref{sec:qcorscope} we describe the QCOR language extension scope. In Sec.~\ref{sec:qcorextensions}  we define a single-source, quantum-classical programming model (Sec.~\ref{sec:programming}), discuss an asynchronous, host-directed execution model (Sec.~\ref{sec:execution}), detail a discrete memory model to go along with the programming model (Sec.~\ref{sec:memory}), and define data structures and associated library calls that form the basis for the asynchronous execution of a quantum-classical hybrid program within a single C or C\texttt{++} source code context. In Sec.~\ref{sec:qcordemo} we illustrate QCOR language with examples. We provide conclusions and outlook in Sec.~\ref{sec:concl}.

\section{Existing Programming Approaches}\label{sec:existingapproaches}
A number of programming mechanisms have been developed for fault-tolerant, as well as for NISQ-era, computers. In this section we describe and categorize related efforts on low-level languages, formal languages, embedded domain specific languages, and frameworks (or software developer kits, SDKs). 

\subsection{Low-level Languages}
Ultimately, all quantum operations are executed via a hardware-level, vendor-specific electronic control system. Often these systems take sequences of digital quantum instructions and transform them into hardware-specific analog control sequences. Recent work is beginning to expose this level of analog programming to users (e.g., OpenPulse \cite{McKay2018}), but to date, such efforts are limited. 
In order for programmers to adequately leverage quantum computing resources, there must exist a low-level interface that maps programs to the quantum device's driver controls in order to enact quantum computation at the physical level. This interface may be considered as a primitive quantum assembly language that higher-level programming tools target in order to interface with vendor-supplied analog driver systems. This assembly language sits at the low-level. It consists of gate-like instructions implementing single or two-qubit unitary operations \cite{Mike&Ike}, but it can also be specified in terms of a broader set of analog attributes \cite{McKay2018}. 

Languages for circuit representation have been provided recently by commercial hardware vendors. For example, OpenQasm has been introduced by researchers at IBM for the expression of experiments in the IBM Quantum Experience, and enables the expression of quantum circuits via the composition of parameterized one-qubit gates and two-qubit entangling gates \cite{Cross2017}. Likewise, Quil, put forward by researchers at Rigetti \cite{Smith2016}, defines an abstract machine model for quantum-classical computations. Both of these languages essentially provide a front-end programming mechanism for IBM/Rigetti quantum computers that interfaces high-level tools (see Sec.~\ref{sec:F&L}) with the vendor-specific hardware. Both also expose syntax for quantum subroutines (compositions of primitive gates) and conditional statements -- i.e., instructions conditioned upon a measurement outcome of another specified qubit. 

\subsection{Formal Languages}
Formal quantum programming language (QPL) development efforts began to arise in the late 1990s and have led to the specification of both imperative and functional 
languages~\cite{Gay:2006:QPL:1166042.1166045}. These formal languages are stand-alone and do not leverage an existing classical host language. Perhaps the best example of a general QPL is the Quantum Computation Language (QCL) put forward by Bernhard Omer  \cite{qcl}. QCL is an imperative, C-like approach that exposes primitive types for quantum data (qubits) and typical constructs for classical control. This approach (and most others like it) serves as a programming language for an ideal quantum device, and does not actively incorporate near-term error mitigation. Moreover, without specification of execution and memory  models, formal languages lack quantum compilation and execution strategies. 

\subsection{Embedded Domain Specific Languages}
Moving toward a higher level of expressiveness, recent efforts have begun to focus on the development of embedded domain specific languages (EDSL) for quantum computation. Microsoft has defined \QS~ a scalable (can be used to program quantum hardware with tens to millions of qubits in theory), multi-paradigm (supports both functional and imperative programming models), domain-specific programming language for quantum computing~\cite{Svore2018}. It enables programming at multiple levels of abstraction, and it can describe instructions that are executable on both simulators and physical hardware. By design, it \QS~does not define quantum states or other properties of quantum mechanics directly, but rather does so indirectly through the action of the various subroutines defined in the language. It \QS~is a strongly typed language and comes with a set of built-in primitive types as well as support for user defined types.

Quipper, developed as part of the IARPA QCS project and released in 2013, is another example of an EDSL~\cite{quipper-intro}. Self-defined as an \emph{embedded functional programming language}, it is built on the Haskell classical functional language and uses an extended circuit model of quantum computation  \cite{quipper}. It can generate representations with trillions of gates  \cite{quipper} and is independent of the backend gate-model quantum hardware~\cite{quipper_new_scientist}. Quipper has an extensible data model that allows programmers to build upon the core types. To perform computations in Quipper, one must write functions, with each function composed of a definition line defining input and output (i.e., ``type'' of the function) followed by the main body defining the quantum operations to be executed in sequential order~ \cite{quipper-intro}.

A further prototypical example of an EDSL is Scaffold, which is a high-level imperative programming language that extends C with data types for quantum and classical bits, as well as functions that act as primitive quantum gates \cite{scaffold}. Scaffold programs are composed of a hierarchical set of quantum modules (sub-circuits) and classical control constructs such as loops and conditionals. The compiler implementation for Scaffold, \texttt{ScaffCC}, modifies the core LLVM intermediate representation to provide a way to parse or compile high-level gate sequences to the LLVM intermediate representation. This enables the compiler (and developers) to inject custom IR passes that perform common quantum compilation tasks, such as multi-qubit gate decomposition, simplifications, and reductions to a targeted gate set. 

\subsection{Frameworks and Libraries}
\label{sec:F&L}
To connect low-level quantum assembly with higher-level languages commonly used by domain scientists, researchers have recently developed software frameworks enabling quantum program expressability via standard object-oriented data structures.
These efforts have resulted in open-source Pythonic frameworks exposing data structures and functions for the composition, compilation, and execution of quantum circuits. To date, the most popular Pythonic approaches are the Qiskit \cite{McKay2018} and pyQuil \cite{Smith2016} frameworks from IBM and Rigetti, respectively.

OpenFermion is an open-source package for computing and manipulating representations of fermionic systems with an emphasis on quantum chemistry \cite{openfermionarxiv}. The core functionality of OpenFermion is the provision of mappings from the space of electronic structure problems to qubit-based quantum representations -- i.e., into representations that can be compiled for and executed on a quantum computer. Included in this package are classes for forming various types of Hamiltonians, including molecular, Fermi-Hubbard, and Bose-Hubbard Hamiltonians. A general FermionOperator class is provided that stores a sum of products of fermionic creation and annihilation operators. After a problem Hamiltonian has been cast in a second quantized representation, it can be mapped to a qubit representation using one of the available spin-fermion transformations such as Jordan-Wigner (JW), Bravyi-Kitaev (BK), and Bravyi-Kitaev super fast (BKSF). 
OpenFermion relies on modular plugins that must be installed seperately for performing classical electronic structure calculations and for simulating and compiling quantum circuits.  Plugins for classical electronic structure calculations include OpenFermion-Psi4 and OpenFermion-PySCF. Circuit compilation and simulation plugins include OpenFermion-Cirq for integration with Cirq, Forest-OpenFermion for integration with Rigetti's Forest, and SFOpenBoson for integration with the Strawberry Fields bosonic computing language \cite{Killoran2019}.

\section{QCOR Language Extension Scope}\label{sec:qcorscope}
The HQCC model has recently emerged as a highly promising avenue for using NISQ devices for a myriad of challenging problems~\cite{farhi2014,tran2016,mcclean2016,Bauer2016,otterbach2017,Li_PRL_2017,Klco2018}. Its viability in the NISQ era is due to (i) lower resource requirements and (ii) resilience against the effects of systematic (unitary) biases. Optimism for the HQCC programming model is fueled by a recent flurry of experimental implementations~\cite{lanyon2010, Peruzzo2014, OMalley2016, Kandala2017, Colless2018, Dumitrescu2018, otterbach2017, wang2017experimental, benedetti2018generative, Paesani2017} coupled with recent HQCC algorithm design progress~\cite{Temme2017,LiSimon2017,McClean2018}.

The HQCC model leverages the power and flexibility of existing classical computing infrastructures but delegates classically intractable tasks to a quantum processor -- i.e., a quantum device may be treated as a co-processor. Classically intractable but quantumly easy function evaluation tasks arise in multiple contexts: for example, optimization of the quantum cost function such as total energy of a many-body quantum system~\cite{Peruzzo2014} over some configuration space or simply implementation of an iterative self-consistent many-body computation~\cite{Bauer2016}.  

An algorithmic example falling within this model of computing is the variational quantum eigensolver (VQE). This algorithm seeks the minimal eigenvalue of some quantum observable $\hat{O}$ by leveraging the variational principle of quantum mechanics. First, a problem-appropriate parameterized circuit realizing a unitary transformation, denoted as $U(\vec{\theta})$, where $\vec{\theta}=(\theta_1,\cdots,\theta_N)$ is the set of parameters, is defined. An initial set of parameters for this circuit is defined and the program is executed on the attached quantum accelerator. Measurements of the quantum state are taken in a manner dictated by the observable of interest, and the composite set of measurements is used to compute $\expect{\hat{O}}_{\vec{\theta}} = \bra{\Psi(\vec{\theta})} \hat{O} \ket{\Psi(\vec{\theta})}$, which is the expectation value of the observable $\hat{O}$ with respect to the parameterized state $\ket{\Psi(\vec{\theta})}$. The result is returned to the classical optimization (or update) routine which updates the parameters as appropriate, and the process is repeated until convergence.

One can extract core abstractions from an algorithm like VQE that are applicable to other HQCC algorithms. The primary concepts at play here are (i) a general quantum observable, (ii) a quantum cost or objective function, (iii) the (parameterized) quantum circuit or kernel, (iv) and the classical optimizer or update rule. The role of the observable is to dictate the measurements (bases) that are required to fully evaluate the quantum function. The objective function uses the observable expectation value to evaluate the quantum cost function and to compute an algorithm-specific task at a given set of parameters. The optimizer or update rule is responsible for iteratively computing the objective function and defining subsequent circuit configurations. We use these core concepts as the seed for the design and specification of a novel language extension for C or C\texttt{++} focused on noisy, quantum-classical, variational computing.

A key consideration for the HQCC paradigm is the noise present in quantum program execution and the need for mechanisms mitigating the effects of noise. Currently available qubits are imperfect and, with physical to logical encoding overheads of $\sim$1000:1, the number of physical qubits required for fault-tolerant schemes is beyond current technological capabilities. Researchers have therefore begun to investigate alternative strategies to improve the accuracy of small-scale quantum programs. These error mitigation strategies often rely on the pre- and post-processing of quantum results. Examples include qubit readout error mitigation \cite{Kandala2017}, zero-limit noise extrapolation \cite{Temme2017,Li2017}, and domain-specific purification schemes \cite{Borle2019}. In addition to the core concepts we distill from variational quantum computing, we further require mechanisms for the integration of custom error mitigation strategies. The goal is to eventually enable the automation of error mitigation as an appropriate API call or ahead-of-time compiler transformation.

\section{QCOR Language Extension}\label{sec:qcorextensions}
\subsection{Programming Model}
\label{sec:programming}

In the HQCC paradigm, a classically hard computational task is broken up into a sequence of quantum and classical subtasks. The former are designed such that a NISQ device can efficiently evaluate them. The latter are designed to be solvable by a classical computer. 
QCOR leverages the HQCC paradigm to develop a programming model that enables a sequence of quantum and classical subtasks to be implemented cooperatively in a single application. The QCOR programming model is a single-source heterogeneous programming model that allows quantum kernels to be implemented within a C or C\texttt{++} application. QCOR's language extensions allow quantum kernels to be constructed using high level C/C\texttt{++} syntax via library calls and directives. The library calls and directives are designed to satisfy the requirements for HQCC algorithms, but could in principle be used for more general schemes, e.g. those involving feed-forward logic. 

\subsection{Memory Model}
\label{sec:memory}
The QCOR memory model assumes that the host memory and quantum control system memory are discrete memory spaces with explicit and implicit memory accesses between the two memory spaces.  
All memory access is explicit on the host device -- i.e., there is a library call or directive expressed in the code by the programmer. On the other hand, memory allocation on and data transfer to the quantum device are implicitly managed by the compiler. For example, an explicit allocation of memory for ``quantum results'' also implicitly requires the same memory for "quantum results" to be accessible on the quantum device. In addition, the host must explicitly allocate memory to create instances of objective functions and observables, and these are implicitly allocated as quantum registers to evaluate the observable on the quantum side. 

\begin{figure*}[ht]
 \centering
  \includegraphics[height=3.5in,width=5.5in]{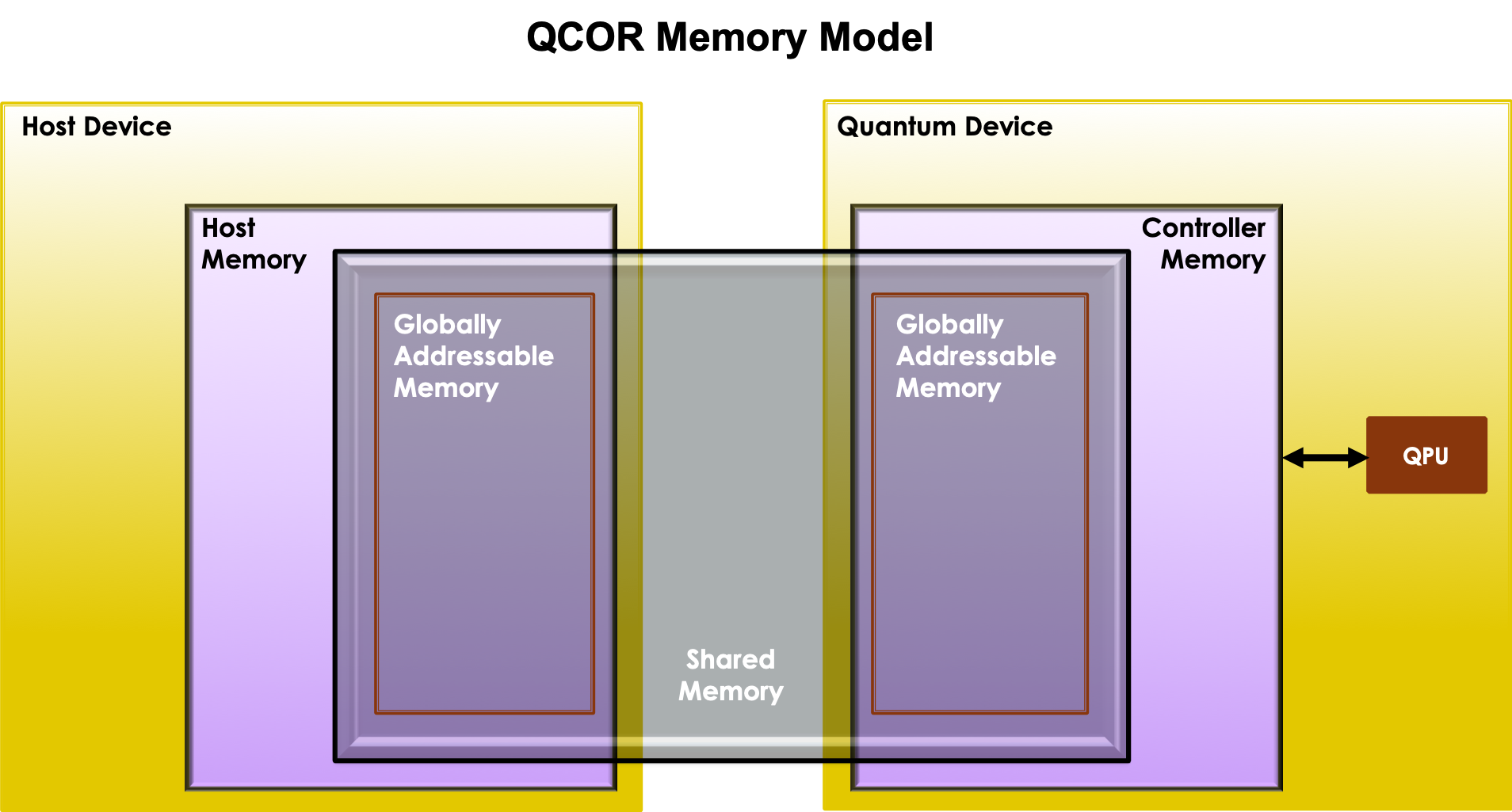}
  \caption{Diagram of the QCOR memory model}
  \label{fig:mem_model}
\end{figure*}

The discrete memory spaces in the QCOR memory model support shared and distributed memory between the host and quantum system. 
When the memory spaces are distributed, a copy of the allocated memory exists in both the host and quantum system address spaces.  
Memory that is written by the quantum system is updated in the host memory address space with library calls on the host. 
Memory written by the host is also updated on the quantum system with library calls on the host.  
When the memory space is shared between the host device and quantum system, we assume that there is an allocatable block of memory where the physical location of the memory is abstracted away, and that a block of memory allocated in this shared space is accessible by both the host and the quantum device. Because this block of memory is accessible by both systems, the library calls that access and update memory do not require a memory copy. Figure \ref{fig:mem_model} provides an illustration of the memory system.

\subsection{Data Structures}
Here we describe the core data structures that are part of the QCOR language extension specification for iterative, quantum-classical application expression. We begin by stating the assumptions underlying our model that language implementors will need to provide or address, and then we move into a detailed description of each of the QCOR data structures. 

\subsubsection{\textbf{Preconditions for Language Implementors}}
The following data structures and associated object model assume a number of preconditions on implementors of this specification. First, the specification assumes that the language being extended provides (through standard or third-party libraries) a heterogeneous associative array that maps string keys to a variant-like value. Our specification refers to this concept as a HeterogeneousMap, in that it provides a way to map string keys to a heterogeneous set of types. Examples of this concept include the \texttt{dictionary} data structure in Python, and a \texttt{map} in C\texttt{++} where the value template type models a \texttt{variant} or \texttt{any} data structure (see C\texttt{++}-17 standard, or third-party approach like \cite{mpark}). We define this data structure in order to promote overall specification maintainability. An example use case for this data structure is in the storage of quantum execution metadata, which is hardware-specific information of various types that is often needed by application programmers for quantum results post-processing, noise mitigation, etc. Specification implementations are free to implement the associative array in a manner appropriate to the native language being extended. Effectively, this array serves as a utility data structure that is available for use within the specification implementation as well as from the public API for use by QCOR programmers. 

\begin{figure}
 \centering
%
\includegraphics[width=5in,height=3.5in]{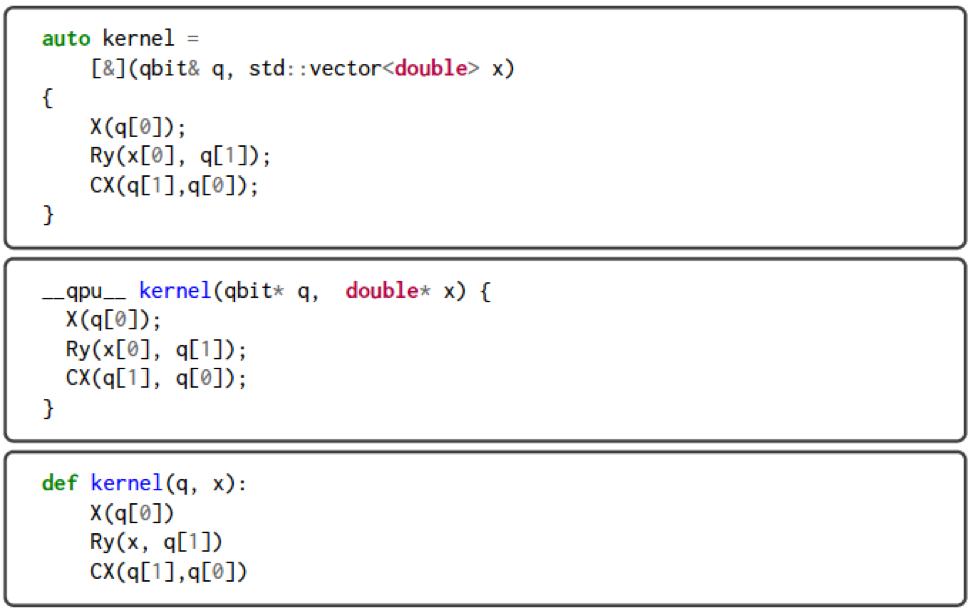}
\caption{Example of valid QCOR quantum kernels in C++ (top), C (middle), and Python (bottom).}
\label{fig:kernels}
\end{figure}

Our specification leaves the exact definition and expression of a quantum code to concrete QCOR language extension implementors. The QCOR data structures and object model denote the concept of a quantum expression as a general \emph{quantum kernel}. Essentially, the kernel represents a composition of individual quantum instructions. The specification speaks in generic terms here in order to be extensible to multiple models of quantum computation and to different expressions of individual quantum instructions. As an example, a prototypical QCOR compiler implementation in C\texttt{++} may choose to represent a quantum kernel as a lambda or functor-like structure whose body contains individual quantum instructions, and in this way promote the single-source nature of the QCOR specification. See Fig. \ref{fig:kernels} for examples of expressing a quantum kernel in C++, C, and Python.
\subsubsection{\textbf{Observable}}
The Observable data structure represents a rule over a range of qubits that defines the measurement bases after the circuit has been executed on the quantum hardware. The measurements return a set of observed bitstrings indicating that qubit \emph{n} has been observed in state \emph{s}, for every qubit within the range. The \texttt{Observable} can be defined as a list of measurement bases, thereby providing a mechanism for taking an unmeasured circuit and producing a list of (one or many) measured circuits. Observables expose an API that enables internal construction from a string representation or from a HeterogeneousMap instance.

The following examples are illustrative of how the Observable can be implemented:
\begin{itemize}
\item \emph{PauliObservable} - representation of a spin Hamiltonian (e.g., $H = X_0X_1 + Z_0Z_1$). The PauliObservable serves as the fundamental or default Observable, as all Observable implementations must eventually be mapped to this form.

\item \emph{FermionObservable} - representation of a fermionic, second quantized Hamiltonian (e.g., $H = c^\dagger_0 c^\dagger_1 c_0 c_1$). The FermionObservable exposes an API that implements pertinent algebraic rules for constructing a complex \texttt{Observable} from simpler Observables of the same type. This type requires delegation to an ObservableTransform sub-type (see Section \ref{sec:obstransform}) to map itself to a PauliObservable. 

\item \emph{ChemistryObservable} - representation of a molecular electronic structure problem as specified by a basis set, geometry (Z-matrix), charge, and other properties. The ChemistryObservable type implements internal construction (e.g., leveraging Libint \cite{Libint2}) from a HeterogeneousMap containing the above properties. The ChemistryObservable serves as a prototypical example of extending the Observable interface for a complex domain-specific Observable generation subsystem. 

\end{itemize}
The Observable abstraction exposes the following methods to be implemented by sub-types:
\begin{itemize}
    \item \textbf{observe} - Map an unmeasured circuit to a list of measured circuits based on this Observable's measurement bases. Takes as input the bare circuit Kernel and outputs a list of locally transformed Kernels. 
    \item \textbf{fromString} - Internally construct this Observable based on a user-provided string. 
    \item \textbf{fromOptions} - Internally construct this Observable based on a user-provided HeterogeneousMap. 
\end{itemize}

\subsubsection{\textbf{ObservableTransform}}
\label{sec:obstransform}
The ObservableTransform serves as an extension point for general transformations of Observables. It exposes a simple API for mapping one Observable to another. The ObservableTransform is needed to map (via a Jordan-Wigner, Bravyi-Kitaev, etc. transformation) a high-level Observable (e.g., field theoretic or fermionic operators) to the computationally fundamental PauliObservable. This concept can be extended to a broader set of transformations that seeks to reduce or optimize the Hamiltonian Observable into a more compact representation, thereby better conforming to the limitations of NISQ hardware.
The ObservableTransform exposes the following methods to be implemented by sub-types:
\begin{itemize}
    \item \textbf{transform} - Takes as input an Observable and returns a generally transformed Observable.
\end{itemize}

\subsubsection{\textbf{ResultBuffer}}
The ResultBuffer corresponding to a single quantum execution provides a container for the quantum state measurement results (i.e., a measurement-to-counts associative array). Furthermore, the ResultBuffer contains a HeterogeneousMap instance that provides a mechanism for keeping track of execution-specific metadata. The ResultBuffer can optionally contain a sub-list of children ResultBuffers, thereby providing an implementation of the familiar composite design pattern (a tree of ResultBuffers) \cite{gof}. This functionality enables the return of a single ResultBuffer that contains the results from all iterations.
The ResultBuffer concept exposes the following attributes:
\begin{itemize}
    \item \textbf{metadata} - A HeterogenousMap containing execution-pertinent information.
    \item \textbf{quantum-results} - A mapping of the measurement result (e.g., as a binary string) to the number of times the result was observed. 
\end{itemize}

\subsubsection{\textbf{ObjectiveFunction}}
The ObjectiveFunction represents a functor-like data structure that models a general parameterized scalar function. It is initialized with the problem-specific Observable and Kernel and exposes a method for evaluation, given a list or array of scalar parameters. Implementations of this concept are problem-specific, and leverage the observe() functionality of the provided Observable to produce one or many measured Kernels that are then queued for execution on the available quantum co-processor, given the current value of the input parameters. The results of these quantum executions are to be used by the ObjectiveFunction to return a list of scalar values, representing the evaluation of the ObjectiveFunction at the given set of input parameters. Furthermore, the ObjectiveFunction has access to a global ResultBuffer that it uses to publish execution results at the current input parameters. 
The ObjectiveFunction concept exposes the following methods to be implemented by sub-types:
\begin{itemize}
    \item \textbf{dimensions} - Returns the number of ObjectiveFunction parameters required. 
    \item \textbf{operator()} - Takes as input the ObjectiveFunction parameters, executes this ObjectiveFunction, and returns the corresponding value of this ObjectiveFunction at those parameters.
\end{itemize}

We also intend for the ObjectiveFunction concept to enable automated error mitigation. We allow this through further extensions of the ObjectiveFunction which, in effect, decorate concrete, algorithmic ObjectiveFunctions. Decoration of ObjectiveFunctions implies that sub-types take as input other algorithmic ObjectiveFunctions and then implement \texttt{operator()} to execute pre-processing, the decorated ObjectiveFunction, and post-processing of execution results, and then return the otherwise error-mitigated result. This pattern enables the general pre- and post-processing steps that are inherent to a number of popular error mitigation strategies on gate model quantum computers. 

\subsubsection{\textbf{Optimizer}}
The Optimizer serves as an extension point for classical multi-dimensional function optimization. This interface exposes an optimize routine that takes an ObjectiveFunction as input and outputs the optimal function parameters and associated function value. Note that pseudo-optimizer functions could also be used to simply provide an update rule for self-consistent calculations, and that by default, if no optimizer is provided, the ObjectiveFunction result is simply returned. 

\subsection{Execution Model and Library Routines}
\label{sec:execution}
\begin{figure*}[ht]
 \centering
 \includegraphics[width=5in,height=4in]{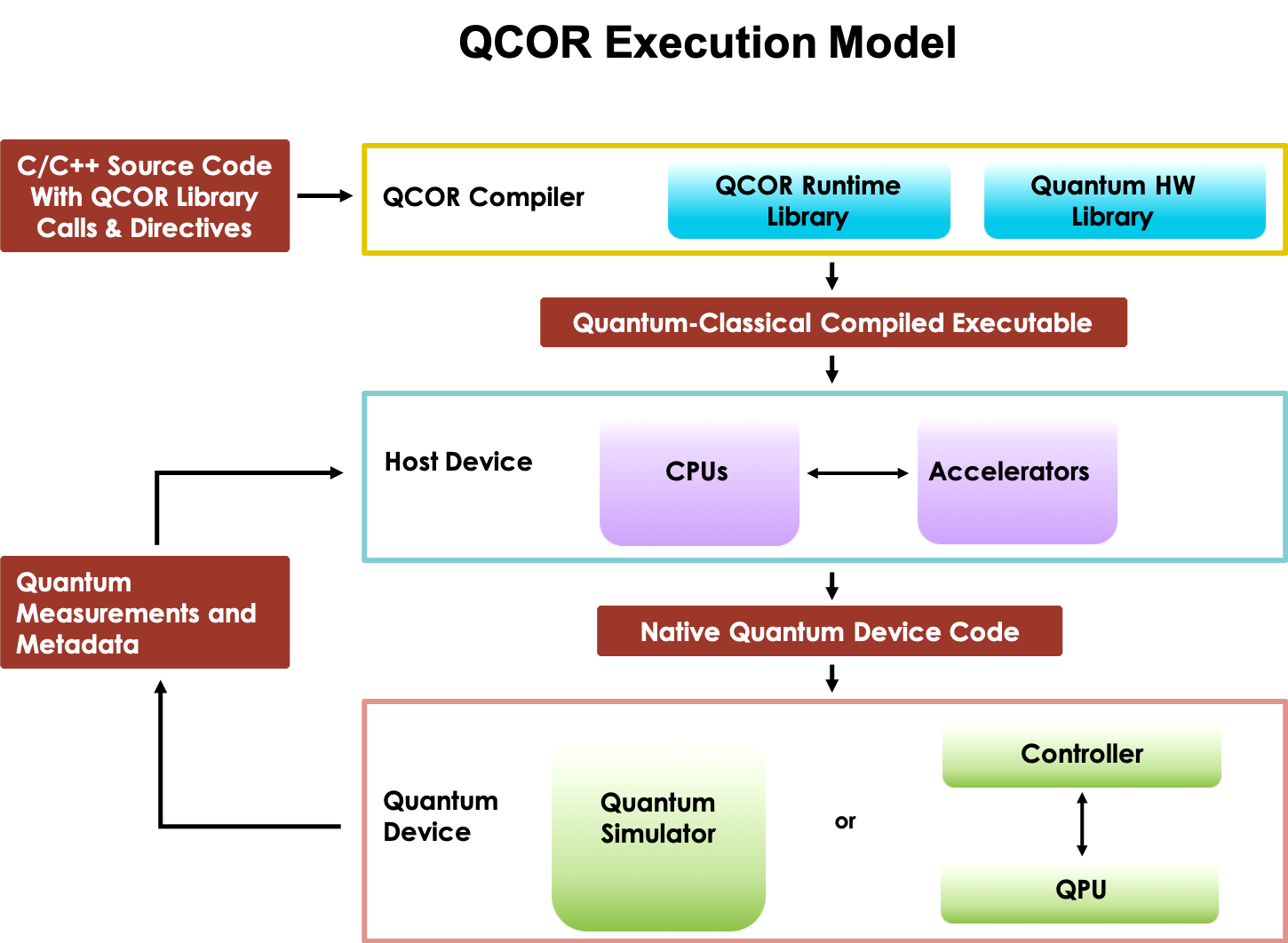}
  \caption{Diagram of QCOR Execution Model}
  \label{fig:exec_model}
\end{figure*}

QCOR targets an execution model whereby application execution is directed by the host with access to an attached quantum device. Applications are broadly thought of as containing two components: a main, classical part, and one or more quantum kernels or subroutines. Figure \ref{fig:exec_model} illustrates QCOR's execution model. Designated quantum kernels are compiled and offloaded to the quantum device. The quantum device executes QCOR library calls and/or QCOR regions identified by directive notations. When the host encounters a quantum kernel, the quantum kernel, in the form of compiled hardware native or simulator 
instructions, is passed to the quantum device controller. Execution on the host is asynchronous to execution on the quantum device. 

QCOR exposes two primary library calls to the user that enable a wide variety of hybrid quantum-classical use cases: the \texttt{taskInitiate} and \texttt{sync} calls. The host launches a task for execution on the quantum device through the \texttt{taskInitiate} call. This call can be executed with references to the Observable data structure, a quantum kernel (circuit), the ObjectiveFunction, and the Optimizer.

The \texttt{taskInitiate} call orchestrates and composes the workflow necessary to initialize the ObjectiveFunction and begin classical optimization of it through the provided Optimizer. The ObjectiveFunction evaluations, driven by the Optimizer, make calls to the quantum device. Immediately upon invocation, this call returns a handle object that programmers use to synchronize the asynchronously executing task. At this point, programmers are free to start other work in parallel to the asynchronously executing tasks. When ready, programmers can invoke the second QCOR library call, which implements thread synchronization. This \texttt{sync} call blocks execution on the host until the terminating condition of the ObjectiveFunction is reached and returns the ResultBuffer containing all quantum execution metadata and measurement results.

The arguments to the \texttt{taskInitiate} call can take on default values, thereby leading to a set of variants of the \texttt{taskInitiate} call itself. Our specification stipulates that if the Observable is absent from this call, then QCOR assumes an Observable that dictates a measurement on all qubits in the computational basis. If the kernel is absent, the call assumes the identity kernel which maps the zero state to the zero state. If the ObjectiveFunction is absent, the call assumes a default objective function that simply returns the expectation value of the given Observable at the given set of parameters. In the absence of an Optimizer, the \texttt{taskInitiate} call requires that users provide a concrete set of ObjectiveFunction parameters and will simply return the evaluation of the ObjectiveFunction at those parameters.

The aforementioned calls specified by QCOR enable several forms of parallelism within a quantum-classical compute context. The three forms of parallelism that QCOR targets are: i) data-level parallelism; ii) quantum bases-level parallelism; iii) task-level parallelism. 
At the data-level, multiple data may be processed in parallel. For example, a quantum cost function may be evaluated at a variety of configurations across distinct QPUs, or within non-overlapping regions of a single QPU, and can be executed by submitting multiple \texttt{taskInitiate} calls in parallel. Bases-level parallelism refers to a purely quantum level of parallelism without a classical counterpart. Objective functions derived from Hamiltonian observables must often be evaluated in a variety of non-commuting bases (distinct measurement settings) in order to evaluate the observable expectation value. Measurement bases can be parallelized by defining and measuring observables that are a subset of the original observable. These basis measurement executions can be parallelized across distinct QPUs or within non-overlapping regions of a single QPU. Lastly, QCOR's execution model is amenable to task-level parallelism using separate \texttt{taskInitiate} calls. 

We leave the exact nature of the implementation of these quantum-classical parallel computing paradigms to concrete QCOR language extensions, but we expect some combination of the data structures and execution model presented here in tandem with a concurrent programming model, such as the Message Passing Interface (MPI).

\section{Demonstration}\label{sec:qcordemo}
\begin{figure}
 \centering
%
%
%
\includegraphics[width=5in,height=3.5in]{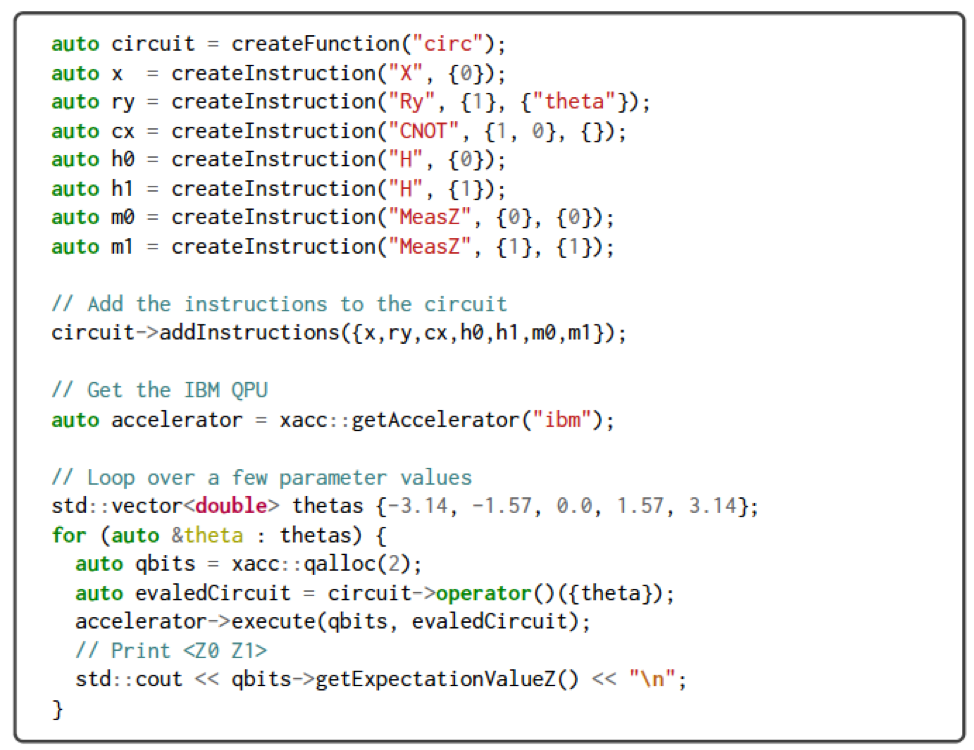}
\caption{Implementation of a simple two-qubit variational quantum eigensolver algorithm using the lowest level of the XACC API. This snippet demonstrates an existing native C\texttt{++} approach to programming hybrid quantum-classical algorithms.}
\label{fig:xaccvqe}
\end{figure}

Here we demonstrate the utility of the QCOR language extension specification. To do so, we illustrate the expressiveness of the language without delving into implementation details. Specifically, we show how to express a set of hybrid quantum-classical and iterative application kernels. We show first how current approaches enable these types of expressions and then move into the demonstration and discussion of our QCOR expressions. 

Existing approaches for quantum programming require users to leverage framework data structures for quantum program expression and backend execution. The QCOR specification seeks to alleviate most of the boilerplate expressions required for programming quantum code and hybrid applications. We start by demonstrating a prototypical example of the variational quantum eigensolver algorithm using the base, lowest-level API from the XACC framework \cite{xacc}. The code snippet shown in Figure \ref{fig:xaccvqe} is prototypical of a number of programming frameworks and approaches. Users begin by constructing quantum programs (circuits) using available quantum instructions (some parameterized by rotation angles) and container circuit data structures. These instructions are added to the circuit container structure, and reference to the IBM quantum backend is requested. Next, users manually loop over concrete rotation angles to trace out the expected value of a given observable (here $X_0 X_1$) as a function of the angle. 

\begin{figure}
 \centering
%
%
%
\includegraphics[width=5in,height=2.5in]{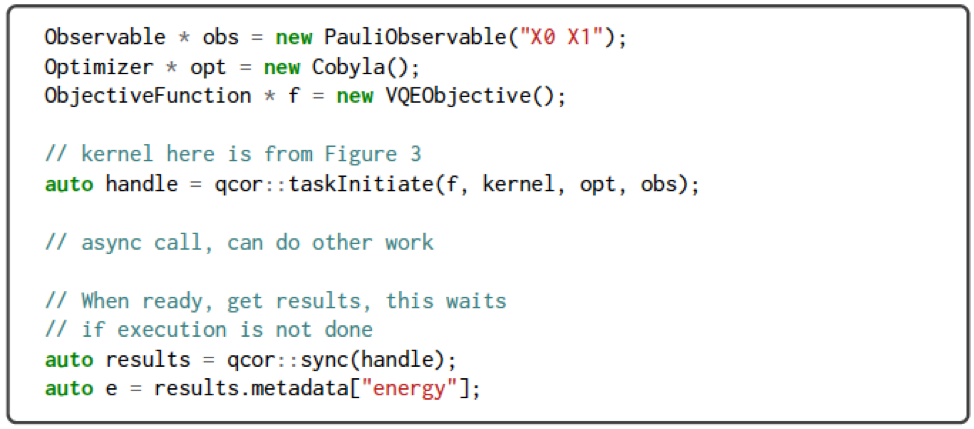}
%
\caption{Prototype implementation of a simple two-qubit variational quantum eigensolver algorithm with the QCOR language extension. Note that \texttt{kernel} here refers to an implementation-specific kernel expression like that in Figure \ref{fig:kernels} (top).}
\label{fig:qcor_example}
\end{figure}

\begin{figure}
 \centering
%
%
%
\includegraphics[width=5in,height=3in]{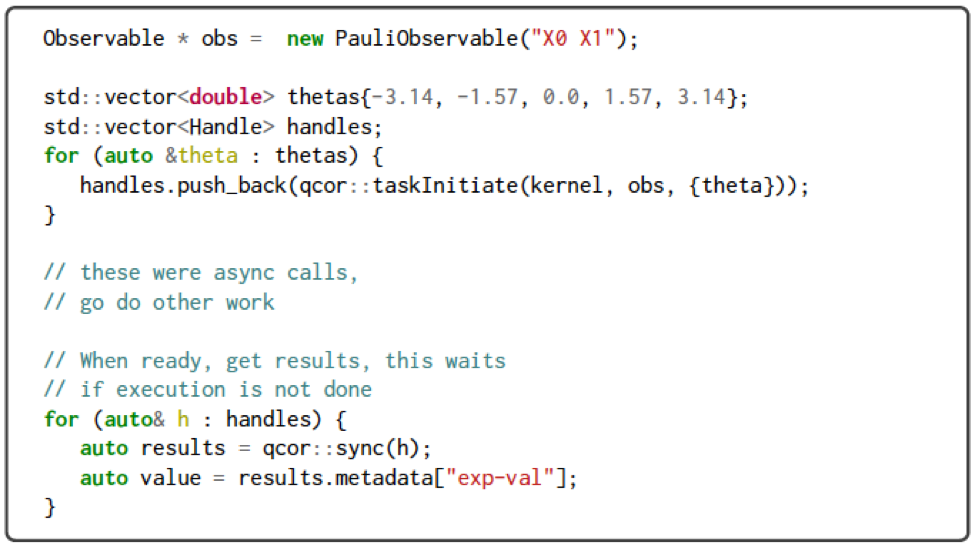}
%
\caption{Prototype implementation demonstrating \texttt{taskInitiate} with default ObjectiveFunction and no Optimizer, taking a vector of kernel parameters.}
\label{fig:qcor_example2}
\end{figure}

QCOR provides a more expressive, higher-level approach for these types of tasks, as demonstrated in Figure \ref{fig:qcor_example}, and does so in the native language of the underlying QCOR implementation. This relieves the need for complex circuit book-keeping data structures, and enables programmers to express quantum code in the native dialect. QCOR operates at the level of Observables, Optimizers, and ObjectiveFunctions, and therefore leaves itself extensible for any number of sub-type implementations. As noted before, Observables can be implemented for domain-specific problems, Optimizers can be implemented for any number of classical function optimization routines, and ObjectiveFunctions can be implemented for any number of domain- and algorithm-specific functional tasks. In Figure \ref{fig:qcor_example}, we demonstrate how the same problem expressed in Figure \ref{fig:xaccvqe} can be expressed in the QCOR language extension specification, through appropriate QCOR concept sub-types. Our language extension specification data structures enable extension to a number of algorithms and applications. In Figure \ref{fig:qcor_example}, we start by defining the Observable, Optimizer, and ObjectiveFunction concepts to provide implementations specific to an $X_0 X_1$ spin Hamiltonian, constrained optimization by the linear approximation (COBYLA) classical optimization routine, and a variational quantum eigensolver objective function, respectively. These instances, and the quantum kernel, are used to seed the QCOR \texttt{taskInitiate} library call, which uses the information available in these types to optimize the ObjectiveFunction, which itself delegates to the execution of measured kernels, dictated by the Observable. Optimization proceeds over a number of iterations, and during this time users are free to do other work due to the default asynchronicity of the \texttt{taskInitiate} call. Using the returned handle reference from \texttt{taskInitiate}, users can call the \texttt{sync} QCOR library call to get the execution results. This call syncs the user thread with the asynchronous thread upon task completion, and returns an instance of the ResultBuffer populated with all execution results. 

\begin{figure}
 \centering
\includegraphics[width=5in,height=0.5in]{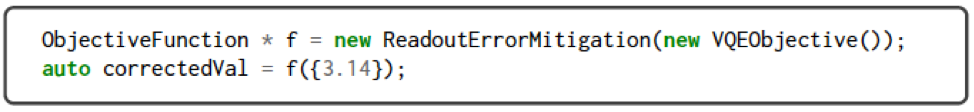}
%
\caption{Prototype implementation demonstrating automated error mitigation through ObjectiveFunction decoration.}
\label{fig:qcor_example3}
\end{figure}

As discussed in Section \ref{sec:execution}, there a number of variants of the \texttt{taskInitiate} method that can be used to compose quantum-classical applications. Figure \ref{fig:qcor_example2} demonstrates this library call with kernel, Observable, and parameter arguments only. This implies that the default ObjectiveFunction will be used which is simply the evaluation of the expectation value $\expect{\hat{O}}_{\vec{\theta}} = \bra{\Psi(\vec{\theta})} \hat{O} \ket{\Psi(\vec{\theta})}$ at the provided $\theta$, where $\hat{O}$ is $X_0 X_1$.

Finally, to enable automated error mitigation, one simply needs to decorate the desired ObjectiveFunction with an ObjectiveFunction implementing the mitigation strategy or strategies. Figure \ref{fig:qcor_example3} demonstrates the composition, or decoration, of ObjectiveFunctions to enable various layers of error mitigation. Here we show a prototypical qubit measurement readout error mitigation strategy implemented as an ObjectiveFunction, delegating to a concrete algorithmic ObjectiveFunction. We note that with this design, any number of error mitigation strategies can be composed.

\section{Conclusion} \label{sec:concl}
We have presented the initial specification of the QCOR language extension for heterogeneous quantum-classical computing. Our specification defines a means for efficient expression of quantum-classical applications in a high-performance computing context. We have defined programming, memory, and execution models that enable a single-source, asynchronous approach to quantum-classical computing, whereby quantum kernels are expressed with high-level language constructs. Our data model provides a robust mechanism for near-term, iterative approaches, and can be leveraged in a post-NISQ fault tolerant setting as well through appropriate compiler extensions provided by implementations. Finally, we have provided prototypical code snippets (based on a concrete implementation of QCOR that is underway) that demonstrate the expressiveness of our specification and its overall efficiency with respect to existing approaches. Moreover, we have described a mechanism for leveraging QCOR to provide programmers with automated error mitigation for near-term, noisy applications. 

\begin{acks}
The authors acknowledge DOE ASCR funding under the Quantum Computing Application Teams program, FWP number ERKJ347. This research used quantum computing system resources supported by the U.S. Department of Energy, Office of Science, Office of Advanced Scientific Computing Research program office.
\end{acks}

\bibliographystyle{ACM-Reference-Format}
\bibliography{main}

\end{document}